\documentclass[aps,prl,showpacs,superscriptaddress,preprint,onecolumn]{revtex4-1}

\usepackage{graphicx}
\usepackage{color}
\usepackage{bm}

\begin{document}

\title{{\Large Single 100-terawatt attosecond X-ray light pulse generation}}
\bigskip
\author{\textbf{X. R. Xu}}
\affiliation{Center for Applied Physics and Technology, HEDPS, State Key Laboratory of Nuclear Physics and Technology, and School of Physics, Peking University, Beijing, 100871, China}
\author{\textbf{B. Qiao}}
\email[Correspondence should be addressed to: ] {bqiao@pku.edu.cn}
\affiliation{Center for Applied Physics and Technology, HEDPS, State Key Laboratory of Nuclear Physics and Technology, and School of Physics, Peking University, Beijing, 100871, China}
\affiliation{Collaborative Innovation Center of Extreme Optics, Shanxi University, Taiyuan, Shanxi 030006, China}
\author{\textbf{Y. X. Zhang}}
\affiliation{Center for Applied Physics and Technology, HEDPS, State Key Laboratory of Nuclear Physics and Technology, and School of Physics, Peking University, Beijing, 100871, China}
\author{\textbf{H. Y. Lu}}
\affiliation{Center for Applied Physics and Technology, HEDPS, State Key Laboratory of Nuclear Physics and Technology, and School of Physics, Peking University, Beijing, 100871, China}
\author{\textbf{H. Zhang}}
\affiliation{Center for Applied Physics and Technology, HEDPS, State Key Laboratory of Nuclear Physics and Technology, and School of Physics, Peking University, Beijing, 100871, China}
\affiliation{Institute of Applied Physics and Computational Mathematics, Beijing 100094, China}
\author{\textbf{B. Dromey}}
\affiliation{Department of Physics and Astronomy, Queen's University Belfast, Belfast BT7 1NN, United Kingdom}
\author{\textbf{S. P. Zhu}}
\affiliation{Institute of Applied Physics and Computational Mathematics, Beijing 100094, China}
\author{\textbf{C. T. Zhou}}
\affiliation{Center for Applied Physics and Technology, HEDPS, State Key Laboratory of Nuclear Physics and Technology, and School of Physics, Peking University, Beijing, 100871, China}
\affiliation{Institute of Applied Physics and Computational Mathematics, Beijing 100094, China}
\author{\textbf{M. Zepf}}
\affiliation{Department of Physics and Astronomy, Queen's University Belfast, Belfast BT7 1NN, United Kingdom}
\affiliation{Helmholtz Institut Jena, Fröbelstieg 3, 07743 Jena, Germany}
\author{\textbf{X. T. He}}
\affiliation{Center for Applied Physics and Technology, HEDPS, State Key Laboratory of Nuclear Physics and Technology, and School of Physics, Peking University, Beijing, 100871, China}
\affiliation{Institute of Applied Physics and Computational Mathematics, Beijing 100094, China}

\begin{abstract}
The birth of attosecond light sources is expected to inspire a breakthrough in ultrafast optics, which may extend human real-time measurement and control techniques into atomic-scale electronic dynamics. For applications, it is essential to obtain a single attosecond pulse of high intensity, large photon energy and short duration. Here we show that single 100-terawatt attosecond X-ray light pulse with intensity ${1\times10^{21}}\textrm{W}/\textrm{cm}^{{ 2}}$ and duration ${7.9} \textrm{as}$ can be produced by intense laser irradiation on a capacitor-nanofoil target composed of two separate nanofoils. In the interaction, a strong electrostatic potential develops between two nanofoils, which drags electrons out of the second foil and piles them up in vacuum, forming an ultradense relativistic electron nanobunch. This nanobunch exists in only half a laser cycle and smears out in others, resulting in coherent synchrotron emission of a single pulse. Such an unprecedentedly giant attosecond X-ray pulse may bring us to view a real attoworld.

\end{abstract}

\maketitle

Generation of short attosecond pulses requires significantly high harmonics \cite{Krausz2009,Krausz2016,Rodel2012} with a broad spectral bandwidth. High harmonic generation (HHG) through relativistic laser interaction with solid targets has been identified as a promising way to get bright attosecond pulses \cite{Gibbon1996,Gordienko2004,Gordienko2005,Baeva2006,Dromey2006}, because solid-density plasmas can withstand larger charge separation field for electron oscillations, resulting in much higher harmonics beyond those in gases \cite{Chini,Wheeler,Popmintchev}. For intense lasers, relativistically oscillating mirror (ROM) mechanism \cite{Baeva2006,Dromey2006} has been shown to support generation of coherent attosecond pulse trains in the extreme ultraviolet (XUV) regime, where HHG is due to periodic Doppler upshifted reflection of the laser from collective oscillations of the step-like critical plasma surface. The ROM harmonic spectrum has a fast decay scaling of intensity on the harmonic order as $I(n)\propto n^{-8/3}$ before a rapid efficiency rollover at frequency $\omega_{rs}\propto \gamma_{max}^{3}$, where $\gamma_{max}$ is the maximum Lorentz factor of electrons.

Recently, however, a more efficient radiation mechanism has been identified in which a compressed dense electron nanobunch with a $\delta$-like peak density distribution and relativistic energy is formed outside solid target surface, resulting in coherent synchrotron emission (CSE) \cite{AnderBrugge2010,Pukhov2010,Dromey2012} of XUV-/X- rays. The reflected radiation is proportional to the time derivative of the transverse current density in the nanobunch, instead of a simple phase modulation of the laser pulse as in ROM. This results in a substantial increase in both radiation energy and harmonic orders beyond those predicted by ROM scaling. Therefore, the CSE spectrum is characterized by a slow scaling of $I(n)\propto n^{-4/3}$ to $n^{-6/5}$, which is much flatter than the 8/3 power of ROM. Also $\omega_{rs}$ here intrinsically depends not only on $\gamma_{max}$, but also on the thickness $\delta$ and the maximum density $n_{max}$ of the nanobunch. 

\begin{figure}
\includegraphics[width=8.6cm]{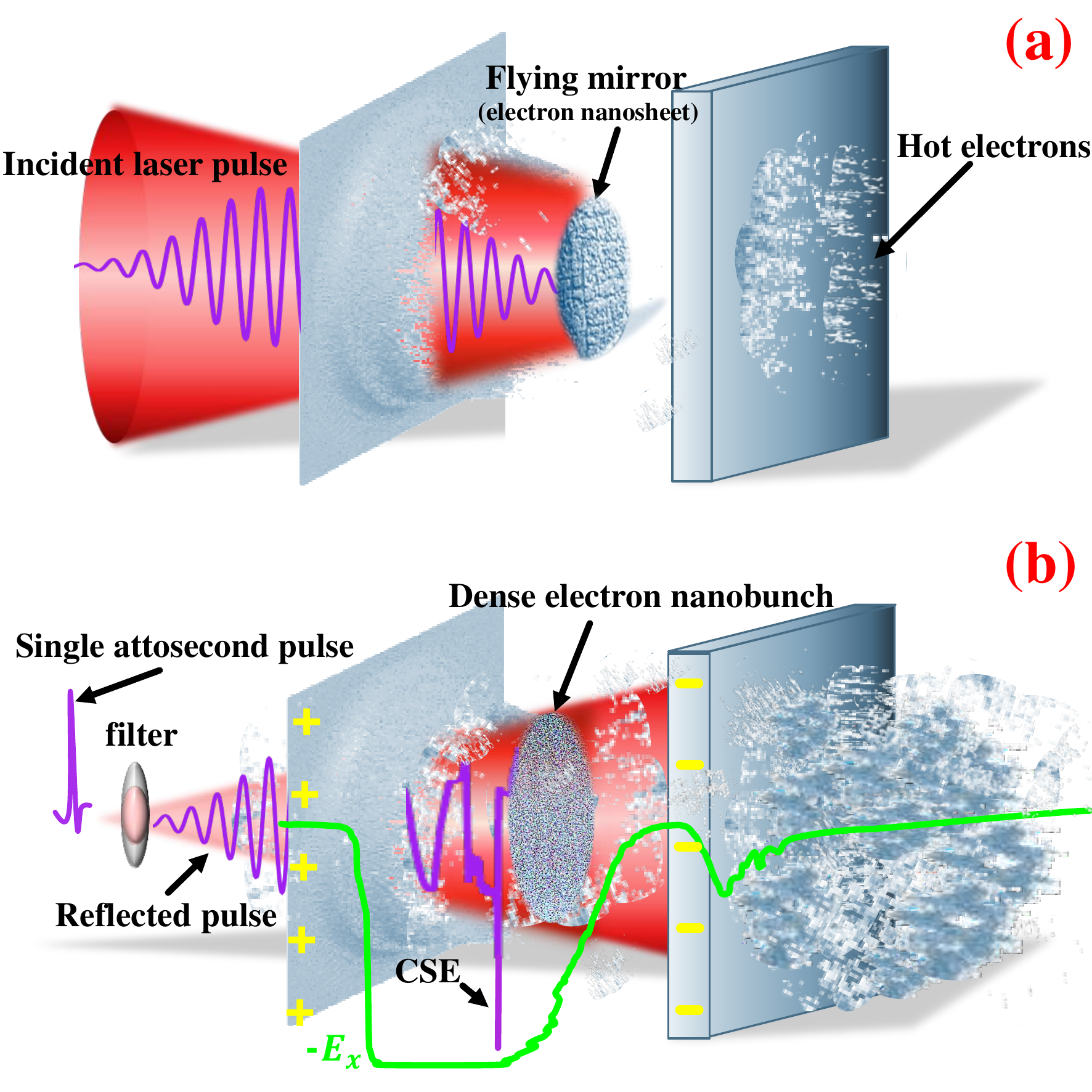}
\caption{{\small {\bf Mechanism of single attosecond X-ray pulse generation by intense laser irradiation on a capacitor-nanofoil target composed of two separate nanofoils.} (a) Electrons of the first nanofoil are blown-out at around the laser cycle of peak intensity, which are rapidly accelerated to relativistic energy, forming a flying dense electron mirror to impact the second nanofoil. (b) The second nanofoil acquires significant negative charges, developing a strong electrostatic potential between two foils. This strong potential drives a large number of electrons out of the second foil and piles them up, forming an ultradense relativistic electron nanobunch for enhanced CSE of single attosecond X-ray pulse.}}\label{fig:schematic}
\end{figure}

CSE presents a promising route to extending attosecond radiations at useful brightness level to the X-ray ($\hbar\omega\sim$keV) regime. However, to obtain a bright single attosecond X-ray pulse required for many applications, unfortunately there are still several key obstacles. First, CSE does not occur in every case of irradiation of intense laser on a solid target, which is highly sensitive to changes in plasma density profile, laser pulse amplitude, pulse duration, angle of incidence, and even the carrier envelope phase of the laser \cite{AnderBrugge2010,Pukhov2010}. The specific conditions required for CSE generation are still unclear. In the specular reflected direction, it is hard to isolate the CSE radiation from the ROM one, where the latter generally dominates. Although the generation of CSE in transmission has been observed \cite{Dromey2012}, the amplitude of the transmitted radiation is much weaker \cite{Dubuis,Cousens}. Secondly, the multi-cycle nature of the drive laser pulses implied that even if the attosecond radiations are emitted by CSE and/or ROM, they must have come in the form of PHz-repetition-rate trains. Although particle-in-cell (PIC) simulations predict that few-cycle, intense laser pulses can be used, however, such laser systems are still under development and most of high power ($>$100-terawatt) laser systems currently in operation in lab deliver pulse with a duration of several 10s femtoseconds.     

To overcome these problems, we propose a new radiation scheme by using a capacitor-nanofoil target irradiated with intense laser pulses, as shown in the schematic Fig. \ref{fig:schematic}, where two nanofoils are separated with a distance less than laser wavelength $\lambda$. The thickness of the first nanofoil satisfies $l_1/\lambda\lesssim(1/2\pi)(n_c/n_1)a_0$ \cite{Qiao2009}, where $a_0$ is the maximum normalized laser amplitude, so that all foil electrons are blown out at around the laser cycle at peak intensity. As shown in Fig. \ref{fig:schematic}(a), the blown-out electrons are rapidly accelerated to relativistic energy by the drive laser, forming a flying dense electron mirror \cite{Qiao2009}, which impacts the second nanofoil quickly. Then, the second foil suddenly acquires significant negative charges, leading to developing of a strong electrostatic potential between two foils. This strong potential difference drives a large number of electrons out of the second foil and piles them up in the vacuum gap, forming an ultradense relativistic electron nanobunch [see Fig. \ref{fig:schematic}(b)]. CSE occurs as the nanobuncn obtaining both the maximum energy $\gamma_x$ and the maximum compressed density $n_e$ in the same peak-intensity cycle, where intense attosecond X-ray pulse is generated. At later cycles, the blown-out relativistic electrons return back from the rear of the second nanofoil and fill in the gap between two foils, smearing out the electrostatic potential, and both foils expand, leading to longer plasma scale length. These break up the conditions for CSE occurrence any more. 

\begin{figure*}
\includegraphics[width=12cm]{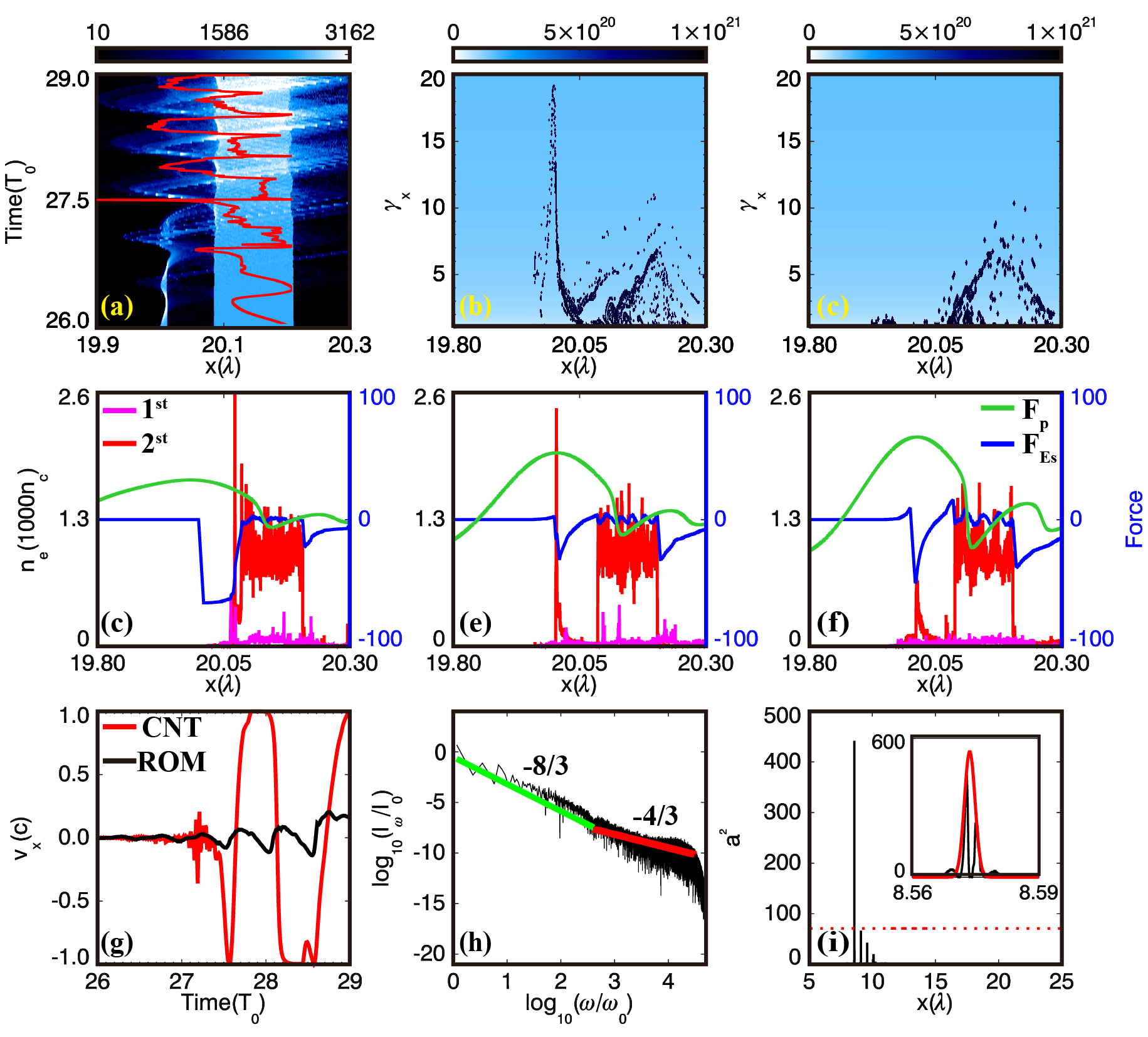}
\caption{{\small {\bf 1D PIC simulation results show the dynamics of coherent single attosecond X-ray pulse generation from the capacitor-nanofoil target.} (a) Electron energy density distribution $n_e\gamma_x$ evolving with time $t$ and the corresponding radiation field amplitude $|E_y|^2$ in the reflected direction (the red line) from $t=26$ to $29.0T_0$. (b) and (c) The electron energy distributions $\gamma_x$ at $t=27.5T_0$ around the peak of the drive laser (where the most intense radiation is emitted) and $28.1T_0$ of the next laser cycle (where the nanobunch significantly spread out), respectively. (d-f) Electron density distributions of the first (pink) and second (red) nanofoils, the electrostatic force (blue), and the laser ponderomotive force (green) at $t=27.48$, $27.57$ and $28.08T_0$, respectively. (g) The velocity $v_x$ of the representative electrons, which contribute to the radiation, for the case of the capacitor-nanofoil target case and the normal single foil target case. (h) Harmonic spectrum for the radiation in the reflected direction, which has a fast decaying power law of -8/3 for $\omega<400\omega_0$ and much flatter law of -4/3 for  $\omega>400\omega_0$ till rollover at around $\omega=20000\omega_0$. (i) The single attosecond intense pulse with peak intensity of $I_{rad}=9.45\times 10^{20}\rm{W}/\rm{cm}^2$, FWHM duration $\tau_{rad}=7.99\rm{as}$ and central wavelength of $\lambda_{rad}=1.0nm$ is obtained, by applying a spectral filter selecting harmonics above $100\omega_0$. Detailed parameters of the simulation are shown in the section of "Methods".}}\label{fig:fig2}
\end{figure*}

In this novel scheme, not only the required stringent condition of CSE can be released, but also the CSE occurrence can be controlled in only one cycle and selected in the specific high-intensity cycle of the drive lasers. More importantly, due to contribution from the electrostatic potential developed in the capacitor-nanofoil target, the formed dense electron nanobunch can be accelerated to larger energy $\gamma_{max}$ and compressed to higher density $n_{max}$ than that in general single foil target case, resulting in much enhanced CSE in the X-ray range. Therefore, an intense single attosecond X-ray pulse can be obtained.

The fundamental dynamics of the proposed scheme is verified using 1D and 2D PIC simulations. Details regarding the simulation setup and parameters are given in the section of "Methods". 1D simulation results are shown in Fig. 2. We see that the radiation process can be divided into three stages. First, at early stage, the rising front of the laser pulse irradiates on the first nanofoil, where the laser intensity is low and the foil keeps opaque. The reflected radiation is dominantly due to ROM from the oscillating surface of the first nanofoil, leading to weak, slightly-compressed radiation pulse, as shown with the portion of the red line from $t=26$ to $27T_0$ in Fig. \ref{fig:fig2}(a). Secondly, when the laser cycle around the peak intensity (from $t=27$ to $28T_0$) starts to irradiate on the first nanofoil, where the laser ponderomotive force is larger than the maximum charge separation field in the foil as $a_0>2\pi(n_1/n_c)(l_1/\lambda)$, almost all electrons are blown-out towards the second nanofoil. The second nanofoil acquires significant negative charges, leading to developing of a strong electrostatic potential between two foils [see the blue line in \ref{fig:fig2}(d)]. When the oscillating laser ponderomotive force decreases, this strong potential drives a large number of electrons out of the second foil and piles them up, forming an ultradense electron nanobunch with relativistic energy. At $t=27.5T_0$, the nanobunch is compressed to the maximum density of $n_{max}\simeq2500n_c$ [\ref{fig:fig2}(e)] and accelerated to the maximum energy of $\gamma_{x,max} = 19.5$ [\ref{fig:fig2}(b)]. Meanwhile, the nanobunch interacts with the peak of laser fields [see the green line in \ref{fig:fig2}(e)], having the maximum transverse current density as well. As a result, a remarkable radiation pulse with extreme broad spectral bandwidth and high amplitude is synchronously emitted, as seen in the sharp peak of the red line at $t=27.5T_0$ in Fig. \ref{fig:fig2}(a). Lastly, afterwards, the blown-out relativistic electrons return back from the rear of the second nanofoil and filled in vacuum. Meanwhile, the electrostatic potential between two nanofoils disappears. Both of these lead to smearing out and decompression of the nanobunch [\ref{fig:fig2}(c) and \ref{fig:fig2}(f)] and eventually break down of CSE in later laser cycles. Fig. \ref{fig:fig2}(g) clearly shows that the electrostatic potential developed in the capacitor-nanofoil target contributes to not only formation of dense electron nanobunches but also significant enhancement in acceleration of electrons to much higher energy than that in the normal case of a single foil target. 

\begin{figure*}
\includegraphics[width=16cm]{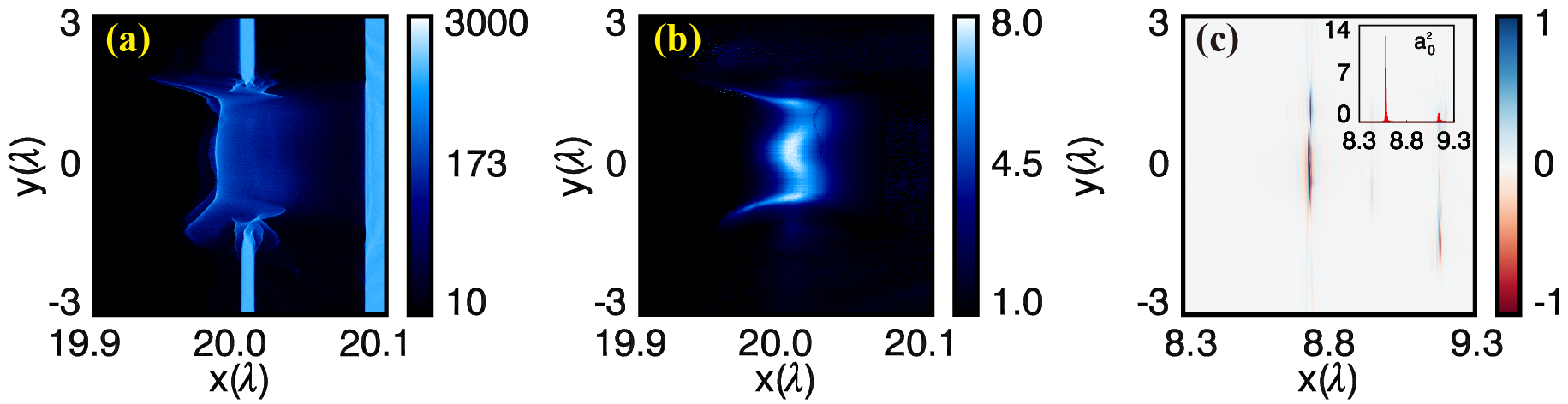}
\caption{{\small {\bf 2D PIC simulation show the scheme is still valid in the multi-dimensional case.} (a) and (b) Electron density and energy ($\gamma_x$) distributions at time $t=27.58T_0$ around the peak of the drive laser, which clearly show that a flat ultradense relativistic electron nanobunch is formed, which contributes to CSE of the second attosecond X-ray pulse. (c) and (d) By selecting harmonics above 100$\omega_0$, a single attosecond X-ray pulse is produced with $I=2.7\times 10^{19} \rm{W}/\rm{cm}^2$ and FWHM duration $\tau=14\rm{as}$. Due to the distortion of the nanobunch under Gaussian laser, the following radiation will transmit into different oblique direction. More details of the simulation setup are shown in the section of "Methods"}}\label{fig:fig3}
\end{figure*}

Figure \ref{fig:fig2}(h) plots the harmonic spectrum of the radiation. In the lower frequency range from $\omega=\omega_0$ to $400\omega_0$, the spectrum shows a power law with the exponent $-8/3$. By contrast in the higher frequency range for $\omega>400\omega_0$, it shows a much more slowly-decaying power law with the exponent $-4/3$, in consistence with the CSE spectral feature. The harmonic spectrum extends up to $\omega=20000\omega_0$, i.e. the photon energy $\sim20\rm{keV}$. Such a broadband spectrum leads to production of extremely short radiation pulse. Applying a high-pass spectral filter selecting harmonics above $100\omega_0$, an intense single attosecond X-ray pulse with peak intensity $I_{rad}=9.45\times 10^{20}\rm{W}/\rm{cm}^2$, FWHM duration $\tau_{rad}=7.99\rm{as}$ and central wavelength of $\lambda_{rad}=2.4nm$ is obtained, as seen in Fig. \ref{fig:fig2}(i). If assuming such a pulse has a transverse radius of $3\mu\textrm{m}$, its power is about 260 terawatts.

\begin{figure*}
\includegraphics[width=10.0cm]{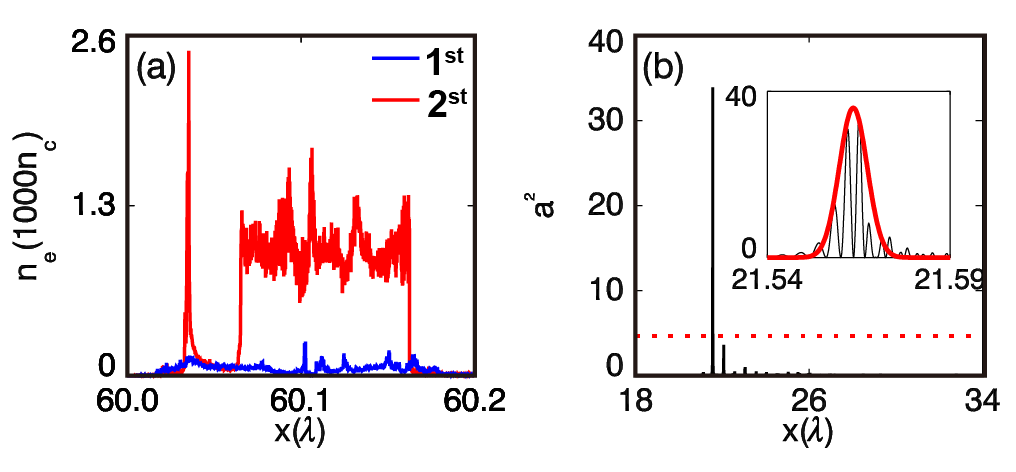}
\caption{(color online) {\small {\bf A demonstrative experiment of the proposed scheme is proposed with state-of-the-art petawatt-femtosecond lasers.} (a) shows the experimental setup. (b) and (c) show the 1D PIC simulation result: A compressed ultradense relativistic electron nanobunch of density $2500n_c$ is formed (b), and a single attosecond X-ray pulse with intensity $I=7.3\times 10^{19}W/cm^2$ and FWHM duration $\tau=24as$ is obtained with the same filter. Here the laser pulse has a FWHM duration of $25fs$ and other parameters are the same as above.}}\label{fig:fig4}
\end{figure*}

To consider the multi-dimensional effects, the results of 2D PIC simulation with coarser resolution are shown in Fig. \ref{fig:fig3}. A dense electron nanobunch with compressed density of $1500n_c$ and relativistic factor $\gamma_x=8.0$ [\ref{fig:fig3}(b)] is also formed in front of the target, which synchronously emits X-ray radiations, although the transversely non-planar intensity distribution and the coarser simulation resolution may both lead to destructive interference. Figures \ref{fig:fig3}(c) and \ref{fig:fig3}(d) show that a single attosecond X-ray pulse at $I=2.7\times 10^{19} \rm{W}/\rm{cm}^2$, $\tau=14\rm{as}$ and central photon energy $>\rm{keV}$ in the normal reflected direction is also obtained by the same spectral filter for $\omega>100\omega_0$.

Finally, as discussed above, the proposed scheme is rather robust, which can be easily testified in experiments with state-of-the-art high-power lasers, in particular, petawatt-femtosecond lasers. To be more close to the currently available laser conditions, we choose the laser FWHM duration as $\tau=25\rm{fs}$, containing $40$ laser cycles, where it is generally thought no CSE can occur and no single attosecond radiation pulse can be obtained. However, the simulation results show that by using the capacitor-nanofoil target, an intense single attosecond X-ray pulse with $I=7\times 10^{19}\rm{W}/\rm{cm}^2$ and $\tau=24\rm{as}$ [Fig.\ref{fig:fig4}(b)] can be produced by CSE.

In summary, we have found a novel, robust scheme to get intense single attosecond pulse in the X-ray regime via the efficient CSE mechanism, which can be achievable by high-power petawatt lasers. By using the capacitor-nanofoil target irradiated with intense laser, a single 100-terawatt attosecond X-ray light pulse with high intensity $10^{21}\textrm{W}/\textrm{cm}^{ 2}$ and short duration $ 7.9 \textrm{as}$ can be obtained in the reflected direction. Such an extreme attosecond light shows a dramatical increase in both the flux and the photon energy available for engineering light forces. Orders of magnitude more powerful than previous reported attosecond pulses will permit attosecond probing of processes with much smaller cross-section and pump-probe spectroscopies of a wide range of bound electron exception and relaxation dynamics, which are inaccessible today. On the other hand, extending attosecond pulse at useful levels to X-ray (keV) energies will allow capturing the picometer-attosecond-scale motions of electrons via attosecond X-ray diffraction.

{\small
\section{Methods}

The relativistic particle-in-cell code, EPOCH \cite{Ridgers,Arber}, is used for 1D and 2D simulations of the proposed scheme. In 1D simulation of Fig. \ref{fig:fig2}, a P-polarized laser pulse with peak intensity of $I_0=7.7\times10^{21}\rm{W}/\rm{cm}^2$ ($a_0=60$) and wavelength $\lambda =800nm$ is used to irradiate on the the capacitor-nanofoil target. The laser pulse has Gaussian profiles in time with full-width-at-half-maximum (FWHM) duration $\tau=4T_0$, where $T_0=2\pi c/\lambda$ is the laser period of one cycle. Both foils of the capacitor-nanofoil target have a steep density profile with initial solid density of $n_1=1050n_c$ and $n_2=1000n_c$, respectively, where $n_c=m\epsilon_0\omega^2/e^2$ is the critical density. The corresponding thicknesses of the foils are respectively $l_1=8\rm{nm}$ and $l_2=100\rm{nm}$, where the first nanofoil is placed between $20.0\lambda \sim 20.01\lambda$ and the second one is located in $20.085\lambda \sim20.21\lambda$, separated by a distance of $d=60\rm{nm}$ in vacuum. The laser starts to interact with the first nanofoil from time $t=20T_0$. The simulation box is 40$\lambda$ long with a resolution of $\lambda/100000$ (0.08nm). 100 particles per cell for target electrons are taken and ions are set to be immobile. 

In the 2D simulation of Fig. \ref{fig:fig3}, the simulation box is composed of $23\lambda \times 10\lambda$. Due to the limitation of computational resources, we reduce the spatial resolution as $\lambda/10000 \times \lambda/1000$ and the number of particles per cell for target to be 8 only. The laser has a supergaussian transverse profile with FWHM radius of $r_0=3.4\lambda$. All other parameters are the same as the 1D simulation above. 

In the 1D simulation of realistic petawatt laser case of Fig \ref{fig:fig4}, the simulation box extends to be $65\lambda$ long, and the capacitor nano-foil target is placed from $60\lambda$ to $60.15\lambda$, where two nanofoils are separated by a distance of $32nm$ with thickness of $8nm$ and $80nm$ respectively. Similarly, due to the limitation of computational resources, we reduce the spatial resolution as $\lambda/10000$.
}

{\small
\section{Acknowledgements}
This work is supported by the National Natural Science Foundation of China, Nos. 11575298, 91230205, 11575031 and 11175026, the National Basic Research 973 Project, Nos. 2013CBA01500 and 2013CB834100, and the National High-Tech 863 Project and the National Science Challenging Program. B.Q. acknowledges the support from Thousand Young Talents Program of China. The computational resources are supported by the Special Program for Applied Research on Super Computation of the NSFC-Guangdong Joint Fund (the second phase).

\section{Author contributions}
B.Q. proposed and was in charge of the research campaign as the principle investigator. X.R.X., B.Q., Y.X.Z. and H.Y.L carried out the simulations, the data analysis and wrote the paper. S.P.Z, C.T.Z and X.T.He contribute significantly to the theoretical interpretation of the simulation results, and improvement of the paper writing. 

\section{Additional information}
Supplementary information is available in the online version of the paper. Reprints and permissions information is available online at www.nature.com/reprints. Correspondence and requests for materials should be addressed to B.Q.
}
\end{document}